\documentclass[twocolumn]{aastex631}

\usepackage{mathtools}
\begin{document}

\title{Infrared spectral fingerprint of neutral and charged endo- and exohedral metallofullerenes}

\correspondingauthor{R. Barzaga}
\email{rbarzaga@iac.es}

\author[0000-0002-9827-2762]{R. Barzaga}
\affiliation{Instituto de Astrof\'{\i}sica de Canarias, C/ Via L\'actea s/n, E-38205 La Laguna, Spain}
\affiliation{Departamento de Astrof\'{\i}sica, Universidad de La Laguna (ULL), E-38206 La Laguna, Spain}
\author[0000-0002-1693-2721]{D. A. Garc\'{\i}a-Hern\'andez}
\affiliation{Instituto de Astrof\'{\i}sica de Canarias, C/ Via L\'actea s/n, E-38205 La Laguna, Spain}
\affiliation{Departamento de Astrof\'{\i}sica, Universidad de La Laguna (ULL), E-38206 La Laguna, Spain}

\author[0000-0001-6253-6343]{S. D\'{\i}az-Tendero}
\affiliation{Departmento de Qu\'{\i}mica, M\'{o}dulo 13, Universidad Aut\'{o}noma de Madrid, 28049 Madrid, Spain}
\affiliation{Institute for Advanced Research in Chemical Science (IAdChem), Universidad Aut\'{o}noma de Madrid, 28049 Madrid, Spain}
\affiliation{Condensed Matter Physics Center (IFIMAC), Universidad Aut\'{o}noma de Madrid, 28049 Madrid, Spain}

\author[0000-0003-3529-0178]{SeyedAbdolreza Sadjadi}
\affiliation{Laboratory for Space Research, Faculty of Science, The University of Hong Kong, Hong Kong (SAR), PR China}

\author[0000-0002-3011-686X]{A. Manchado}
\affiliation{Instituto de Astrof\'{\i}sica de Canarias, C/ Via L\'actea s/n, E-38205 La Laguna, Spain}
\affiliation{Departamento de Astrof\'{\i}sica, Universidad de La Laguna (ULL), E-38206 La Laguna, Spain}
\affiliation{Consejo Superior de Investigaciones Cient\'{\i}ficas (CSIC), Spain}

\author[0000-0002-3753-5215]{M. Alcami}
\affiliation{Departmento de Qu\'{\i}mica, M\'{o}dulo 13, Universidad Aut\'{o}noma de Madrid, 28049 Madrid, Spain}
\affiliation{Institute for Advanced Research in Chemical Science (IAdChem), Universidad Aut\'{o}noma de Madrid, 28049 Madrid, Spain}
\affiliation{Instituto Madrile\~no de Estudios Avanzados en Nanociencia (IMDEA-Nano), Campus de Cantoblanco, Madrid 28049, Spain}

\author[0000-0002-3938-4211]{M. A. G\'omez-Mu\~{n}oz}
\affiliation{Instituto de Astrof\'{\i}sica de Canarias, C/ Via L\'actea s/n, E-38205 La Laguna, Spain}
\affiliation{Departamento de Astrof\'{\i}sica, Universidad de La Laguna (ULL), E-38206 La Laguna, Spain}

\author[0000-0002-5523-2568]{T. Huertas-Rold\'an}
\affiliation{Instituto de Astrof\'{\i}sica de Canarias, C/ Via L\'actea s/n, E-38205 La Laguna, Spain}
\affiliation{Departamento de Astrof\'{\i}sica, Universidad de La Laguna (ULL), E-38206 La Laguna, Spain}



\begin{abstract}

Small metal-containing molecules have been detected and recognized as one of the hybrid species efficiently formed in space; especially in the circumstellar envelopes of evolved stars. It has been predicted also that more complex hybrid species like those formed by metals and fullerenes (metallofullerenes) could be present in such circumstellar environments.  Recently, quantum-chemical simulations of metallofullerenes have shown that they are potential emitters contributing to the observed mid-IR spectra in the fullerene-rich circumstellar environments of different types of evolved stars. Here we present the individual simulated mid-IR ($\sim$5$-$50 $\mu$m) spectra of twenty-eight metallofullerene species; both neutral and charged endo- and exohedral metallofullerenes for seven different metals (Li, Na, K, Ca, Mg, Ti, and Fe) have been considered. The changes induced by the metal-C$_{60}$ interaction on the intensity and position of the spectral features are highlighted using charge density difference maps and electron density partitioning. Our calculations identify the fundamental IR spectral regions where, depending on the metal binding nature, there should be a major spectral contribution from each of the metallofullerenes. The metallofullerenes IR spectra are made publicly available to the astronomical community, especially James Webb Space Telescope users, for comparisons that could eventually lead to the detection of these species in space.

\end{abstract}



\section{Introduction} \label{sec:intro}
Nowadays, more than 240 molecules (formed by up to 19 different chemical elements) have been detected in space \citep[see e.g.][for a recent review]{McGuire2022}. Different individual molecular species are detected in a variety of cosmic environments, from interstellar/circumstellar media to protoplanetary disks and exoplanet atmospheres and distant galaxies, among others \citep{McGuire2022}. The great majority ($\sim$70\%) of them are organic (carbon-containing) species, mainly detected in the insterstellar and circumstellar medium. Interestingly, metal-containing organic molecules (i.e., with metals like Mg, Na, Al, K, Fe, and Ca) have only been detected in the circumstellar envelopes of evolved stars; especially towards the nearby and bright (prototype) C-rich evolved star IRC$+$10216 \citep[e.g.,][but see also \cite{cernicharo23} for the most recent detections of metal-containing organic species towards IRC$+$10216]{cernicharo87}.

The strong mass loss experienced by low- and intermediate-mass ($\sim1-8$ M$_\odot$) evolved stars during the asymptotic giant branch (AGB) stage provokes the enrichment of the surrounding circumstellar and interstellar medium with the heavier elements (metals and even heavier s-process elements) previously produced inside the star by stellar nucleosynthesis \citep{Karakas2014}. The subsequent post-AGB evolution towards the formation of planetary nebulae (PNe), however, occurs with almost no mass loss neither additional stellar nucleosynthesis \citep[e.g.,][and references therein]{jones2017,kwitter2022}. A metal enrichment of the circumstellar environment not only occurs for PNe precursors but also in other types of astrophysical objects like e.g., the chemically peculiar R Coronae Boraelis (RCB) stars \citep[e.g.,][]{montiel2018,schwab2019,pandey2021}. In these environments, chemical processes take place to form both simple and more complex organic species like fullerenes \citep{Cami2010, Garcia2010, Garcia2011}, but the formation of other organic species containing alkaline to transition metals is also possible.

Thus, the coexistence of organic species and metals in the circumstellar environment of evolved stars can give rise to more complex metal-organic compounds; i.e., hybrid species containing both. The detection of metal-containing organic molecules like MgNC, NaCN, among others, in the circumstellar envelopes around AGB stars and PNe are undeniable examples \citep[e.g.,][]{highberger2001,Ziurys2006,McGuire2022}. Previous experimental studies supported by quantum-chemistry calculations have demonstrated also that fullerenes can react with metals forming endo- and exohedral metallofullerenes \citep[e.g.,][]{Dunk2013}. This has a significant relevance because fullerenes, mainly C$_{60}$ which is characterized by its four strongest mid-IR features at $\sim7.0, 8.5, 17.4$, and 18.9 $\mu$m, have been detected in young PNe and RCB stars \citep[see][]{Cami2010,Garcia2010,Garcia2011}. PNe and RCB stars also have abundant metals residing in their circumstellar envelopes, being thus reasonable to assume that metals would react with fullerenes to produce metallofullerenes.

In order to explore the possible existence of metallofullerenes in fullerene-rich circumstellar envelopes, we have previously performed quantum-chemistry calculations of the IR vibrational spectra of charged and neutral endo- and exohedral metallofullerenes \citep{barzaga2023}. According to the IR spectra simulation \citep[i.e., the total and total-weighted metallofullerene spectra, see][]{barzaga2023} the presence of metallofullerenes could explain the large range of C$_{60}$ 17.4$\mu$m/18.9$\mu$m band ratios observed in PNe and RCB stars. Therefore, metallofullerenes can be considered as potential emitters contributing to the observed IR spectra in the fullerene-rich circumstellar envelopes of these kind of objects.

In this work we focus on the individual simulated mid-IR ($\sim$5$-$50 $\mu$m) spectra of the 28 metallofullerene species; i.e., neutral and charged endo- and exohedral metallofullerenes for seven different metals like Li, Na, K, Ca, Mg, Ti, and Fe. These metals were selected because they are the most abundant and/or well-known to exist in the interstellar medium. According to nucleosynthesis models and published abundances they are also among the most abundant metals in low-mass evolved stars such as PNe and RCB stars \citep[see e.g.,][]{Karakas2014,Jeffery2011}, and thus more likely to react with fullerenes \citep[see also][for more details]{barzaga2023}. We analyze the charge redistribution on the C$_{60}$ carbon cage induced by the metal along with the carbon-metal bond strength and bonding sites, in order to rationalize the changes in the IR intensity and position of the spectral features. To this end, charge density difference mapping and electron density partitioning have been performed. Both methods allowed us to identify the metal-C$_{60}$ interaction effect on the intensity/position of the IR features for the several metallofullerenes. We aim to provide the metallofullerenes IR spectra to the astrophysical community, in particular to the James Webb Space Telescope users, for future spectral match that could eventually lead to the first detection of these complex metal-organic species in space.
\begin{figure*}
\centering
\includegraphics{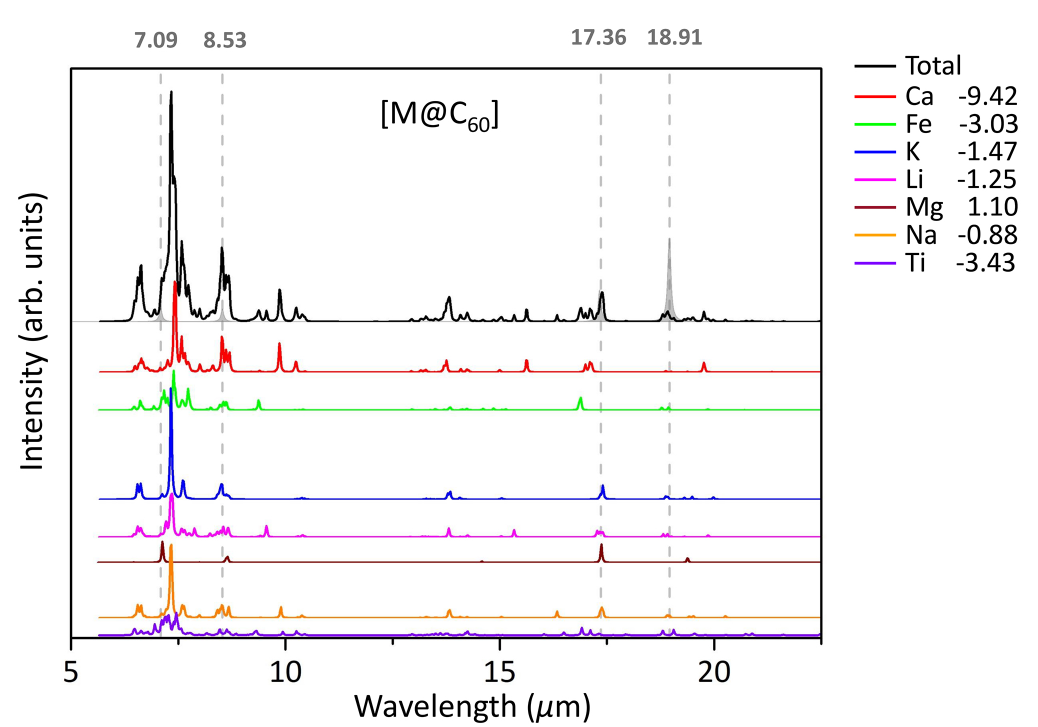}
\caption{Total (black) and individual spectra of neutral metalloendofullerenes [M@C$_{60}$] in the 5-25 $\mu$m range for 7 different metals (Li, Na, K, Ca, Mg, Ti, and Fe). In the case of Ti ([Ti@C$_{60}$]) an intensity offset has been applied to ease the visualization. The legend plot shows the color code used to represent the individual neutral metalloendofullerene spectra. The gray dashed vertical lines and shaded peaks indicate the four strongest IR features of pristine C$_{60}$ at the DFT B3LYP/6-31G level of theory. In addition, binding energies (eV) concerning the metal-C$_{60}$ bond has been also added for each metal. The full range (5-50 $\mu$m) mid-IR spectra can be found in the corresponding appended data.
\label{fig:f1}}
\end{figure*}

\section{Computational Details} \label{sec:comput}
Calculations of the neutral/charged endo (inside C$_{60}$) and exohedral (outside C$_{60}$) metallofullerenes have been performed in the framework of Density Functional Theory (DFT) at the B3LYP/6-31G(d) level \citep{Becke1983,Ditchfield1971}, with the Gaussian 16 code \citep{g16}, following the settings reported in our previous work \citep{barzaga2023}\footnote{In particular, the metals were placed in different binding sites of the C$_{60}$ cage before geometry optimization: hexagon-hexagon bond region, center of a hexagon, direct above a carbon, pentagon-hexagon bond, center of a pentagon and center of cage. After geometry optimization the most stable structure has been selected for each type of metallofullerene (see Table 1). In addition, for all the calculations spin multiplicity (2S + 1) has been considered up to a maximum of S = 5/2.}. The mid-IR spectra were obtained under the harmonic oscillator approximation and, subsequently, wavelengths were adjusted using double scaling factors to account for anharmonicity, vibro-rotational coupling, etc. The maximum error obtained under this approach in the calculated wavelength was $<$ 2\%; previous work have demonstrated the accuracy of this procedure \citep{Robledo2014,Trujillo2022}. For an easier comparison of the several simulated mid-IR spectra, the peak profile has been modelled by a Lorentzian function of FWHM = 0.02 $\mu$m, which reproduces the average resolution ($R\sim$1700) of the Mid-IR Instrument (MIRI; $\sim$5$-$30 $\mu$m) onboard the James Webb Space Telescope (JWST) (but see also below). A Lorentzian function is a more adequate profile of the peak in order to relate molecular structure with IR spectral signature because it describes a peak width invariant around a central wavelength (as expected in gas-phase species) and it reflects the natural width (natural damping and pressure broadening) \citep[see e.g.,][]{pradhan2011}. Also, the former property of the Lorentzian function has been extensively used to reproduce the observed spectral profiles of some Unidentified Infrared bands (UIRs) and Diffuse Interstellar Bands (DIBs) \citep[e.g.,][]{Snow2002,pech2002}. All individual simulated spectra are annexed in the corresponding external data. In any case, the intensity and wavelength of the IR lines are also provided in this paper to allow the astrophysical community the convolution with any peak function; e.g., to obtain simulated spectra at different resolutions and spectral profiles.

Quantum Theory of Atoms in Molecules (QTAIM), employing the AimAll software has been applied to characterize the bonding type and strength by using the topological analysis of the electron density distribution \citep{Bader1990,AimAll}. The former analysis permits the extraction of the kinetic ($G$), potential ($V$) and total $(H = G + V)$ energy densities, with $V$ and $G$ been always negative and positive, respectively \citep{Bader1990,cremer84}. By employing the $|V|/G$ ratio it is possible to determine the bonding type according to: $|V|/G < 1$ indicates a pure ionic bond, while $|V|/G > 2$ corresponds to a pure covalent bond and $1 >|V|/G > 2$ indicates a bond in between \citep{Zabardasti2017,barzaga2021}. The bond strength is identified by the amount of electron density ($\rho$) in the internuclear region of a bond. It is important to note that the topological analysis of the electron density allows an automatic identification of the bonding nature between atoms. Thus, it is possible to determine direct metal-fullerene connections. Finally, charge density difference (CDD) maps were represented in order to illustrate the charge reordering induced by the metal on the carbon cage. The CDD mapping describes the regions within the electron density that experiences an increase (accumulation) or decrease (depletion) of electrons upon metal-fullerene interaction.

\begin{figure*}
\centering
\includegraphics{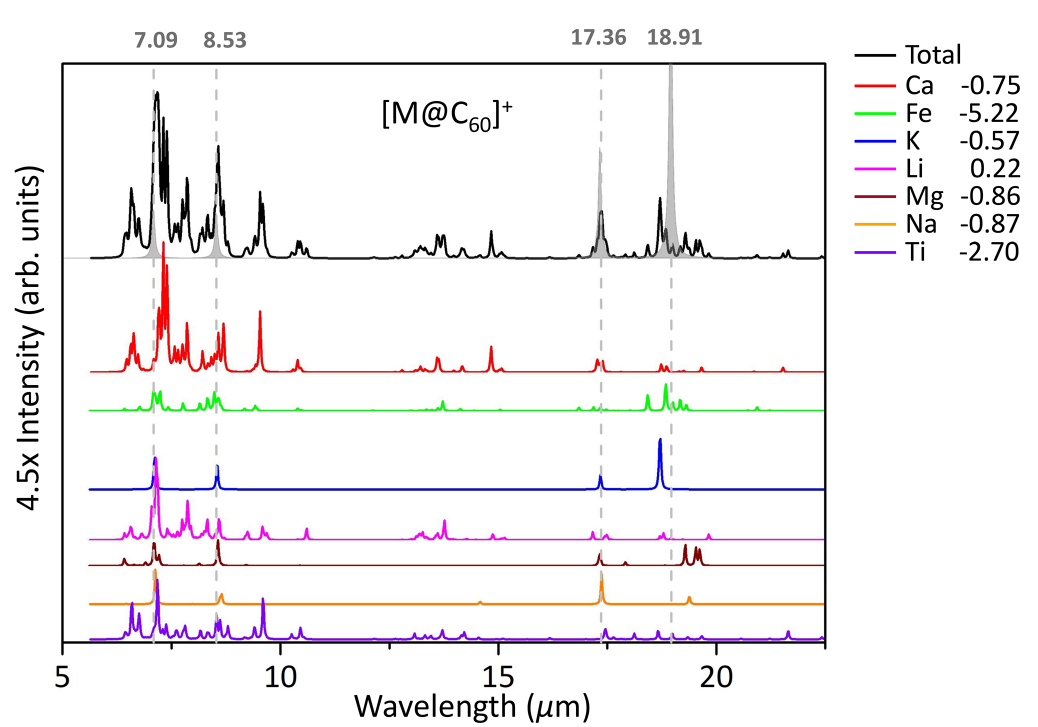}
\caption{Total (black) and individual spectra of charged metalloendofullerenes [M@C$_{60}]^{+}$ in the 5-25 $\mu$m range for 7 different metals (Li, Na, K, Ca, Mg, Ti, and Fe). As in Fig.1, the legend plot shows the color code used to represent the individual charged metalloendofullerene spectra, and the binding energies (eV), while the pristine C$_{60}$ features are marked with gray vertical lines and shaded peaks. Note that this time the intensities are multiplied by factor of 4.5 in order to match with Figure \ref{fig:f1}. Again, the full range (5-50 $\mu$m) mid-IR spectra can be found in the corresponding appended data.
\label{fig:f2}}
\end{figure*}

\section{Results} \label{sec:results}
First, we consider the analysis of the individual mid-IR spectra for the different metallofullerenes, tracking the particular spectral features that define the metal-C$_{60}$ cage interaction for each case. For an easier comparison, Figs. 1 to 4 (representing all individual neutral and charged endo- and exohedral metallofullerenes spectra) also include the total (summed) spectrum for each metallofullerene species as constructed similarly to our previous work\citep{barzaga2023}. The IR intensities of the individual simulated spectra are scaled to the same total value. This way we guarantee that all spectra are at the same intensity level; despite of been vertically offset for a better visualization. In addition, in order to preserve the intensity for each metallofullerene species: exo or endofullerene, the spectra were multiplied also by a scaling factor. In Figs. 1 to 4 we display the spectral range from 5 to 25 $\mu$m, while the full range (5$-$50 $\mu$m) IR spectra can be found in the appended data annexed to this paper. The four strongest features of pristine C$_{60}$ have been also overplotted in Figs. 1 to 4 for comparison purposes and to represent the real difference in their IR intensity against the metallofullerene species. In addition, the binding energies (in eV) are also provided in the figures to show the strength of metal-C$_{60}$ bond; with negative values reflecting a stronger binding. Subsequently, after the spectral analysis, a throughout understanding of the nature of metal-C$_{60}$ bond is carried out in order to recognize it as ionic, covalent or in between. The former is also used to relate the spectral fingerprint of metallofullerenes with its molecular structure.

\subsection{Metalloendofullerenes}
The individual neutral endofullerene $[\rm M@C_{60}]$ IR spectra are shown in Figure \ref{fig:f1}, highlighting the several metal species with different colors. There are two specific regions in these spectra that contains the four pristine C$_{60}$ IR features \citep[see e.g.,][]{Kern2013,barzaga2023}: (i) the 5-10 $\mu$m C-C stretching vibrations; and (ii) the 15-20 $\mu$m C$_{60}$ cage vibrations. The C-C stretching region shows a common trend for Li, Na, and K (alkaline metals) neutral endofullerenes. Multiple features can be distinguished in this region with an intense peak at 7.34 $\mu$m accompanied by a prominent shoulder like in the case of Li. These three ``alkaline'' endofullerenes exhibit different peaks in the 15-17 $\mu$m spectral region; although they display a unique feature centred at $\sim$13.82 $\mu$m. The difference between Li, Na, and K is noticeable in terms of features intensity in the 5-10 $\mu$m region, especially around 7.3 $\mu$m with K showing the most intense feature but less rich spectra (see Figure \ref{fig:f1} and appended data). The next sorted group of neutral endofullerenes are Fe and Ti, which exhibits IR features with similar positioning on the C-C stretching region of the spectra, but not comparable in terms of intensity. The main characteristic of Fe and Ti neutral endofullerenes is the low features intensity across the full spectral region. On the other hand, the Ca neutral endofullerene shows the richer IR spectrum among all neutral metalloendofullerenes; its IR spectrum resemblances the one of others species in some regions but it is not possible to find a very similar spectrum among the rest of endofullerene spectra. Interestingly, the IR spectrum of the Mg neutral endofullerene remains almost identical to the one of pristine C$_{60}$ without any indication of additional emission features in both the C-C stretching and C$_{60}$ cage vibration regions. In general all neutral endofullerenes show a significant loss of features intensity around 18.5-19 $\mu$m, which is the region of the highest intensity features for pristine C$_{60}$. Finally, the contribution to the Total (summed) neutral metallonendofullerene spectrum (black line in Figure \ref{fig:f1}) is dominated by Ca $>$ K $>$ Li in that order, with Mg being the smallest contributor.

In the case of charged metalloendofullerenes $[\rm M@C_{60}]^{+}$, the contribution trend in the mid-IR spectra is hindered by a decrease of the global features intensity with respect to the neutral ones (see the scaling factor used in Figure \ref{fig:f2}). The richer IR spectra observed here correspond to the charged endofullerenes with Ca, Li, Ti, Fe, and Mg. From this group, the Ca charged endofullerene exhibits the highest features intensity, remaining as the main contributor to the Total (summed) charged metallofullerene spectrum (black line in Figure \ref{fig:f2}). Again, the 5-10 $\mu$m C-C stretching region is the one where the spectral modifications are more noticeable, indicating the effect of the metal binding to the carbon cage. In contrast, the Na and K charged endofullerene spectra, when compared with pristine C$_{60}$, display almost unperturbed spectra, which in principle, could be an indication of a weaker metal-carbon cage binding. However, the binding energy values displayed in Figure \ref{fig:f2} do not reflect such as a possible trend because Li is the weakest bonded metal. Thus, another intrinsic effect should be playing a role here as it is discussed further below. From Figure \ref{fig:f2} we can also observe a lack of trend on the spectra landscape of the charged metalloendofullerenes, which indicates different effects of the metal-C$_{60}$ interactions. For example, the Na and K charged endofullerene IR spectra are the most similar ones according to the features position; for the rest of charged endofullerenes, however, there is a diversity of IR bands that are specific to each species. Such observations indicate a significant change in the nature and strength of the metal-C$_{60}$ interaction, depending on the metal, when the molecular system is charged. Consequently, the Total (summed) spectrum (black line Figure \ref{fig:f2}) becomes spectrally broader, weaker, and richer since less species contribute to the same spectral region.
\begin{figure*}
\centering
\includegraphics{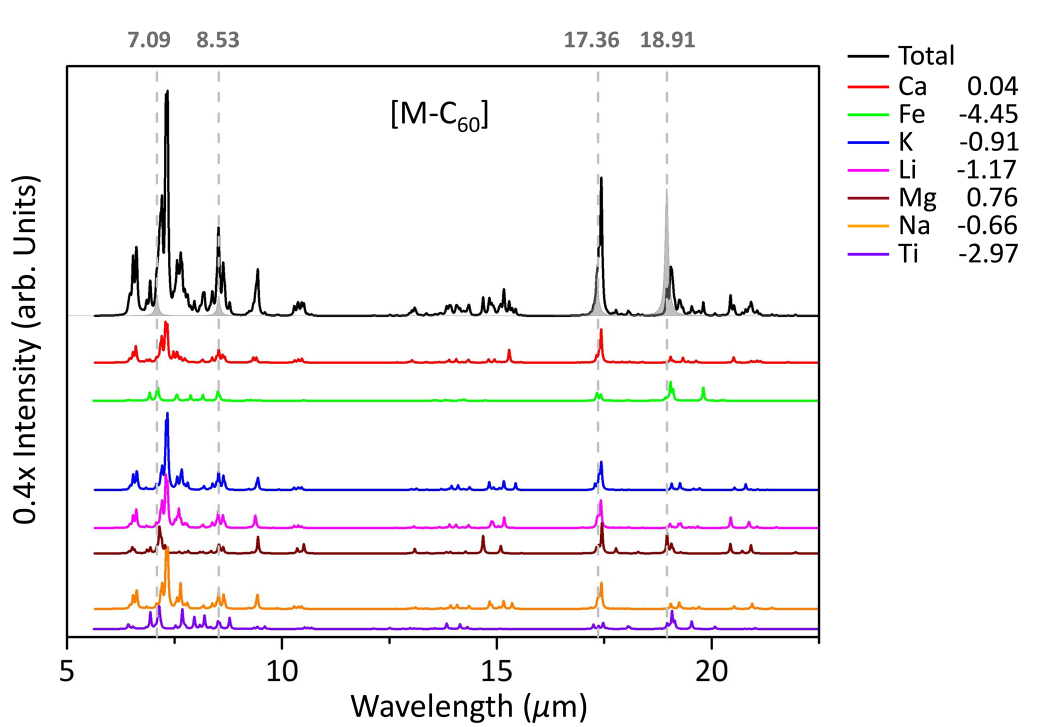}
\caption{Total (black) and individual spectra of neutral metalloexofullerenes [M-C$_{60}$] in the 5-25 $\mu$m range for 7 different metals (Li, Na, K, Ca, Mg, Ti, and Fe). As in Fig.1, the legend plot shows the color code used to represent the individual neutral metalloexofullerene spectra, and the binding energies (eV), while the pristine C$_{60}$ features are marked with gray vertical lines and shaded peaks. Note that this time the intensities are multiplied by factor of 0.4 (i.e., the feature intensities are higher than in Figure \ref{fig:f1}) in order to match with Figure \ref{fig:f1}. Again, the full range (5-50 $\mu$m) mid-IR spectra can be found in the corresponding appended data.
\label{fig:f3}}
\end{figure*}

\subsection{Metalloexofullerenes}
The metalloexofullerenes have the advantage that the metal binds externally to the carbon cage, allowing to accommodate it and establishing a more efficient and stronger interaction than in their endohedral analogues. The C-C stretching region (5-10 $\mu$m) in each metalloexofullerene spectrum reflects such behaviour with multiple IR bands of medium or high intensity. In particular, the IR spectra of neutral exofullerenes $[\rm M-C_{60}]$ (see Figure \ref{fig:f3}) with K, Li, Na, and seemingly Ca, are the most similar ones. Their spectra are characterized by a double peak at 7.26 $\mu$m of different intensity, which is surrounded by medium-intensity and relatively broad satellite bands. Another unique feature of these species is observed at 9.37-9.43 $\mu$m as an isolated band, while a single band at 17.42 $\mu$m (coincident with pristine C$_{60}$) always appears in the C$_{60}$ cage vibration region (15-20 $\mu$m). Both features are also present for the Mg neutral exofullerene but the rest of the spectrum is very different. Finally, the Fe and Ti neutral exofullerene IR spectra form a separate group with an almost full coincidence (both in intensity and position) in their spectral features and a distinctive broad band at 19.05 $\mu$m. It is worth noting that all neutral exofullerene IR spectra exhibit a higher intensity than their endohedral counterparts (see the scaling factors used in Figures \ref{fig:f1} and \ref{fig:f3}), which reinforce the importance of charge rearrangement processes. The metal is externally bonded to the carbon cage, creating a system where charge transfer takes place C$_{60}^{-}$M$^{+}$\citep[see e.g.,][]{barzaga2023}, and thus resulting in a strong modification of the dipole moment and, consequently, of the IR intensity.

\begin{figure*}
\centering
\includegraphics{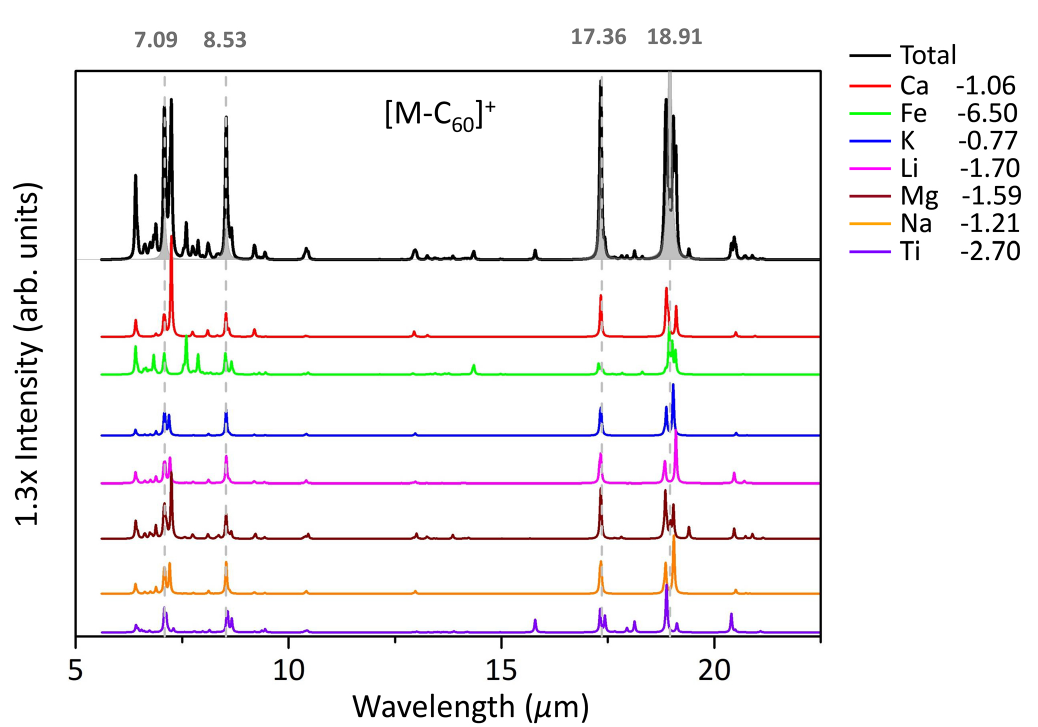}
\caption{Total (black) and individual spectra of charged metalloexofullerenes [M-C$_{60}]^{+}$ in the 5-25 $\mu$m range for 7 different metals (Li, Na, K, Ca, Mg, Ti, and Fe). As in Fig.1, the legend plot shows the color code used to represent the individual charged metalloexofullerene spectra, and the binding energies (eV), while the pristine C$_{60}$ features are marked with gray vertical lines and shaded peaks. Note that this time the intensities are multiplied by factor of 1.3 in order to match with Figure \ref{fig:f1}. Again, the full range (5-50 $\mu$m) mid-IR spectra can be found in the corresponding appended data.
\label{fig:f4}}
\end{figure*}

As expected, charged metalloexofullerenes display a less intense IR spectra (see Figure \ref{fig:f4}) but with spectral features lying very close to those of the pristine C$_{60}$; as calculated by us at the same theoretical level (i.e., at 7.09, 8.53, 17.36, 18.91 $\mu$m). In this case, the charge redistribution induced by the cation creates a system like C$_{60}^{0}$M$^{+}$\citep[see e.g.,][]{barzaga2023}, which likely reduces the changes in the dipole moment, and thus produces a lower IR features intensity. In general, all charged metalloexofullerenes show strong similarities in their spectra, with the exception of the Fe and Ti ones. The IR simulated spectra are characterized by a single band at 6.41 $\mu$m, and a doublet centered at $\sim$7.13 $\mu$m, which can be symmetric for K, Na and Li, and asymmetric for Ca and Mg. Such a doublet seems to be absent for the Fe and Ti charged exofullerene species, which otherwise display an extra band around $\sim$8.53 $\mu$m (i.e., coincident with pristine C$_{60}$). It is to be noted here that only the 6.41 $\mu$m band is free from any contribution from the pristine C$_{60}$ features. At longer wavelengths (i.e., 15-20 $\mu$m), the charged exofullerenes display a very similar spectral behaviour; a band at 17.33 $\mu$m and a doublet centered at 18.98 $\mu$m, both contributing to the pristine C$_{60}$ IR features. Curiously, the Fe and Ti charged exofullerenes are exceptions in this region, showing broader features and richer spectra, respectively.

\subsection{Bonding nature}

The basics of IR spectroscopy are determined by two equations: one obtained from the harmonic oscillator approximation $\left(\bar{\nu} = \frac{1}{2\pi c}\sqrt{\frac{k}{u}}\right)$\footnote{This expression is already modified to determine the wavenumber ($\bar{\nu}$) of the vibration by adding the speed of light constant ($c$). The corresponding conversion factor should be applied in order to obtain wavelength units. In addition, $u$ here represents the reduced mass, and this notation has been used to avoid a misunderstanding with the dipole moment ($\mu$).}, which describes the position (wavelength) of the spectral features. Here the key factor is the force constant $k$ related to the modification of the bond strength. The second expression affecting IR spectroscopy denotes the dependence of the IR intensity with the dipole moment change ($\partial\mu$) as provoked by the vibration displacement ($\partial x$) $I\propto\frac{\partial\mu}{\partial x}$. A modification of the dipole moment can be induced by a strong reordering of the electron density charge, but also by larger vibrational displacements, which hence produce large dipole moment, and thus more intensity. Therefore, in order to make a more exhaustive analysis of the simulated metallofullerene spectra and to infer the origin of the most characteristic features, we have studied the strength and type of the metal-C$_{60}$ bonding as well as mapped the charge reordering in the cage induced by the metal.

\begin{figure*}
\begin{minipage}{0.45\textwidth}
\centering
\includegraphics{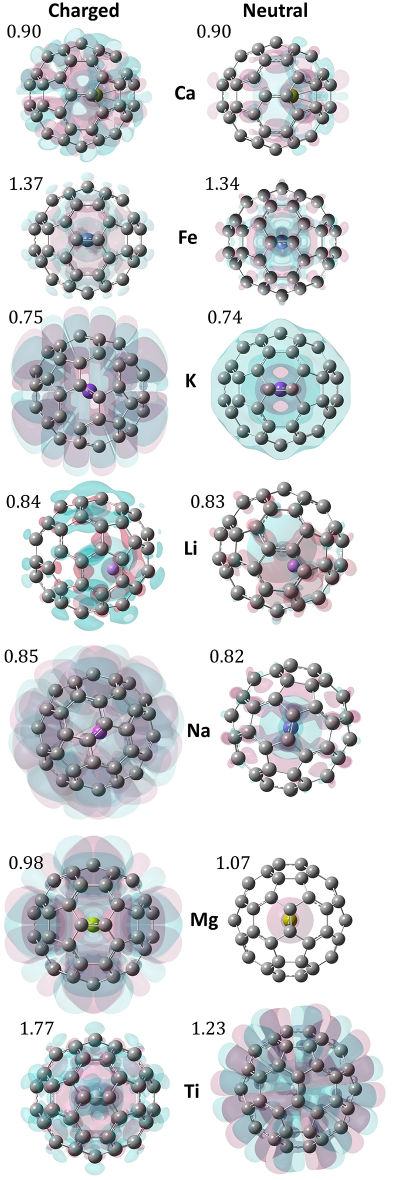}
\caption{Charge density difference maps of the simulated metalloendofullerenes obtained for an isovalue of 0.02 a.u. Light-red regions indicate electrons depletion while light-blue electrons accumulation. In each case the $|V|/G$ ratio is displayed in the top-left corner of the model.
\label{fig:f5}}
\end{minipage}
\begin{minipage}{0.45\textwidth}
 \centering
\includegraphics{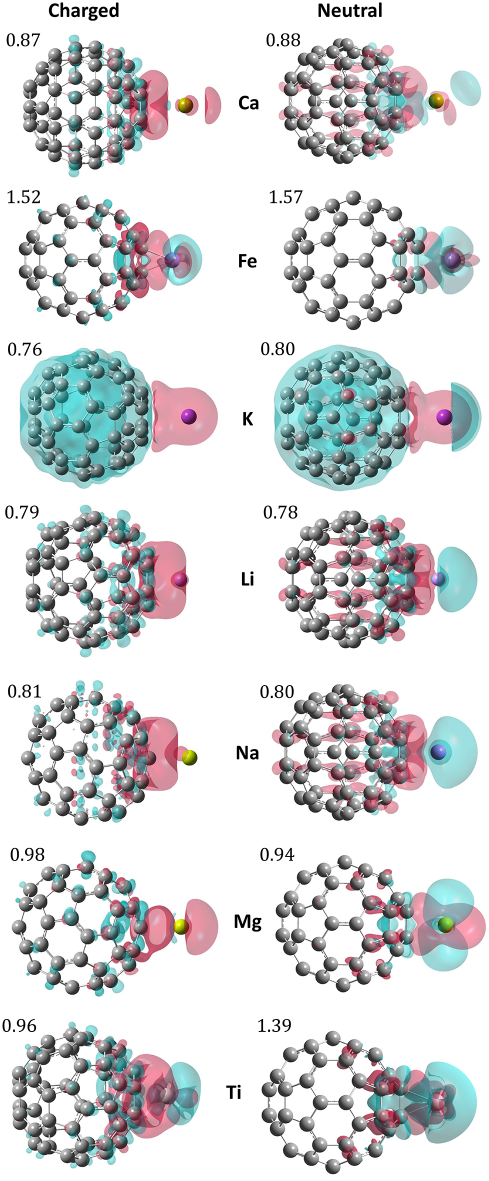}
\caption{Charge density difference maps of the simulated metalloexofullerenes. The notation is the same as in Figure \ref{fig:f5}.
\\\textcolor[rgb]{1.00,1.00,1.00}{12}
\\\textcolor[rgb]{1.00,1.00,1.00}{12}
\\\textcolor[rgb]{1.00,1.00,1.00}{12}
\label{fig:f6}}
\end{minipage}
\end{figure*}

%
%
Figures \ref{fig:f5} and \ref{fig:f6} represent the charge reordering  induced in the carbon cage by the presence of the metal for the endo and exo cases, respectively. Those regions where more accumulation of electrons appear due to the metal presence are depicted in light-blue, while the light-red area encloses those regions with a depletion of electrons. From a first inspection, a higher IR intensity of the exofullerene spectra, in comparison to their endofullerene counterparts, is clear. In contrast to the endofullerenes, the exofullerenes have the metal outside the cage, which creates a larger dipole moment change due to the vibration, increasing the IR intensity. In addition, the charge reordering in the exofullerene creates a significant dipole moment change in comparison to their charged analogous, which is provoked by a depletion and accumulation of electrons at the border of the carbon cage. This behaviour is marked for the neutral exofullerenes with K, Na, and Li; something that explains the large IR intensity observed in Figure \ref{fig:f3}. The neutrality in the endofullerenes shows the same behaviour but at a lower extend. The former is caused by a smaller dipole moment as a consequence of the charge reordering implying the whole carbon cage. Thus, on average, the dipole moment is of lower magnitude. It is worth noting that the metal in neutral endofullerenes have smaller vibrational displacements that also contribute to reduce the IR intensity. Such intensity reduction is stronger when a larger charge reordering takes place, diminishing the dipole moment; as in the case of charged endofullerenes (see the intensity scales in Figs. \ref{fig:f1} and \ref{fig:f2}). Another aspect reinforcing our analysis is the bonding type of the several metallofullerenes as described by the $|V|/G$ ratio (see Section \ref{sec:comput}) and represented in Figures \ref{fig:f5} and \ref{fig:f6}. Predominantly, the metallofullerenes exhibit ionic bondings ($|V|/G < 1$), indicating that the electrostatic interactions dominate, and enhancing large dipole moment changes. In those cases where the ionic character of the bonds is weakened (Fe, Mg, Ti $|V|/G \geq 1$), the IR intensity drastically drops in comparison to the rest of metallofullerenes. This is especially happening for Fe and Ti metallofullerenes, which exhibit bonds closer to covalent ($1 < |V|/G < 2$). Seemingly, the best combination for intense IR spectra is obtained for those systems with a strong charge reordering, large vibrational displacement of the metal, and noticeable ionic bondings.

\startlongtable
\begin{deluxetable*}{cccccccccccccc}
\tabletypesize{\small}
\tablewidth{0.5pt}
\tablecaption{Properties of the metal-C$_{60}$ bonding in metallofullerenes: Average of electron density within the internuclear region ($\rho$) in atomic units (a.u.), number of sites to which the metal is bonded (B) and the position of the metal in the carbon cage. \label{tab:tab1}}
\tablehead{ & \multicolumn{6}{c}{Endofullerenes} && \multicolumn{6}{c}{Exofullerenes}  \\
& \multicolumn{3}{c}{Neutral}& \multicolumn{3}{c}{Charged} &&  \multicolumn{3}{c}{Neutral} & \multicolumn{3}{c}{Charged}\\\cline{2-7}\cline{9-14}
\colhead{} & \colhead{$\rho$} & \colhead{B} & \colhead{Position} & \colhead{$\rho$} & \colhead{B} &\colhead{Position} && \colhead{$\rho$}  & \colhead{B} & \colhead{Position} & \colhead{$\rho$} & \colhead{B} & \colhead{Position}}
\startdata
{  } Ca & 2.27 & 2  & HH     & 2.18 & 2  & HH  && 1.70 & 2 & Hex-As & 1.27 & 5 & Pent\\
{  } Mg & 0.72 & 37 & Cage   & 1.92 & 1  & HH  && 2.96 & 2 & HH & 2.45 & 1 & PH\\
{  } Li & 2.15 & 1  & Hex    & 2.06 & 3  & Hex && 1.79 & 1 & Hex-As & 1.30 & 3 & Hex\\
{  } Na & 1.17 & 1  & HH     & 1.09 & 2  & PH   && 1.40 & 2 & Hex-As & 1.01 & 3 & Hex\\
{  } K  & 0.39 & 34 & Cage   & 0.40 & 60 & Cage && 1.16 & 2 & Hex & 0.78 & 3 & Hex\\
{  } Fe & 8.46 & 2  & HH     & 8.49 & 2  & HH && 8.86 & 2 & HH & 9.97 & 2 & HH\\
{  } Ti & 8.84 & 2  & Hex-As & 24.77 & 2 & HH && 9.47 & 2 & HH & 3.94 & 6 & Hex\\
\hline
\enddata
\tablecomments{\textbf{HH}: the metal is connected in a hexagon-hexagon bond region \textbf{Hex}: center of an hexagon
\textbf{Hex-As}: slightly offset from the hexagon center \textbf{PH}: pentagon-hexagon bond \textbf{Pent}: center of a pentagon \textbf{Cage}: center of C$_{60}$}
\end{deluxetable*}

On the other hand, the binding of the metal to the carbon cage can generate a displacement of the pristine C$_{60}$ features and the appearance of new lines. Two main factors seems to be responsible of the former trend: the position at which the metal is attached and the strength of the bonds. Strong bonds are characterized by the increment of electron density within the internuclear region while the position determines the number of carbon bonds been affected by the metal. Both factors can produce richer IR spectra around the four fundamental C$_{60}$ IR features. An important point here is that the binding energy values displayed in Figs. 1-4 reflect the binding strength (with negative values reflecting a stronger binding) but these values result from the contribution of both metal position and electron density in the internuclear region. For the former it is very difficult to relate the binding energies per se with any possible trend observed in the simulated spectra. Table \ref{tab:tab1} displays the average internuclear electron density ($\rho$), the number of binding sites of the metal (B) and the position to which the metal is bonded. These parameters can be used to understand the peculiarities observed in the metallofullerene IR spectra with the metals Fe and Ti, but also other general trends. It is important to clarify that strong binding does not necessarily imply more intense IR spectra; only just richer IR spectra because the IR intensity mainly depends on charge reordering and metal vibrational displacement (see above). The metals Fe and Ti are by far those more strongly bound to the C$_{60}$ cage (see $\rho$ in Table \ref{tab:tab1}). In particular, the charged metalloendofullerene with Ti, with a $\rho$ = 24.77 a.u. Despite that Ti is bonded to only two carbon atoms at the hexagon-asymmetric position (Hex-As), such interaction is strong enough to affect the other far neighbour atoms and C-C bonds; the latter is translated into a much richer IR spectra (see Figure \ref{fig:f2}). Furthermore, this richness of the spectra is also attained when the metal is at the hexagon-hexagon position (HH) (see the Fe case in Table \ref{tab:tab1}). However, the lowest $\rho$ values displayed in Table \ref{tab:tab1} are observed for both (neutral and charged) endofullerenes with K at the cage position, exhibiting also the larger number of binding sites. This is the reason of the very discrete (or less richer) IR spectra displayed for the K endofullerenes in Figures \ref{fig:f1} and \ref{fig:f2}. The effect of the binding position of the metal seems predominant in the resulting IR spectra, specially when the metals are at similar positions; e.g., the neutral exofullerenes with Li, Na, K, and Ca display some resemblance in their spectra (see Figure \ref{fig:f3}). The general trend observed from the comparison of Table \ref{tab:tab1} with Figures 1-4 is that metals bonded to hexagon-hexagon (HH) sites tend to exhibit richer spectra. In fact, it is possible to establish the following decreasing order according to the metal position: HH $>$ Hex $\simeq$ Hex-As $>$ Pent $>$ PH  $>$ Cage (see Table \ref{tab:tab1}). Therefore, we conclude that the factors to obtain very diverse IR spectra (i.e., in terms of spectral features variety) are the bonding strength and the closeness of the metal to the HH position. Of course, the bonding strength, metal position and charge reordering should be phenomena intrinsically connected but we can try to decoupled their effect in order to understand and predict the IR spectra of new species.

\section{Summary and Astrophysical Relevance}
Quantum-chemistry calculations at the DFT/B3LYP 6-31G(d) level of theory have been performed in order to obtain the individual simulated mid-IR ($\sim$5$-$50 $\mu$m) spectra of 28 metallofullerene species; neutral and charged endo- and exohedral metallofullerenes for seven different metals (Li, Na, K, Ca, Mg, Ti, and Fe)

Our calculations and IR simulations show that metallofullerenes can exhibit rich or more discrete IR spectra regardless of being neutral or charged. Neutral metallofullerenes have the tendency to show richer spectra in the 5-10 $\mu$m spectral region accompanied by a considerable loss of IR intensity in the 15-20 $\mu$m range. Charged metallofullerenes, on the other hand, can display a major IR contribution in the 15-20 $\mu$m range. According to our calculations, however, there are some metallofullerene systems where the neutral species display a discrete spectrum over the full IR wavelength regime. Three key factors determine the landscape of the spectra: charge reordering, bonding strength and metal position. The first one is related to the intensity observed in the IR spectra, while the rest two affect the wavelength displacement and spectral appearance of new IR features.

We conclude that richer spectral regions are not solely describing the presence of neutral species, for instance like aromatic and aliphatic carbon molecules, but also charged species with a strong binding as it is the case of some of the metallofullerenes presented here. Nevertheless, it is very important to stress that for those neutral metallofullerene species with strong charge reordering processes we expect higher IR intensity in comparison to their charged counterparts.

The James Webb Space Telescope (JWST), with a much higher sensitivity (and spectral resolution) than the previous {\it Spitzer Space Telescope}, has the needed potential to unambiguously detect the spectral signatures of new IR emission features potentially due to metallofullerenes (or even other complex metal-organic species). Thus, the metallofullerenes IR spectra are made publicly available to the astrophysical community, especially to the JWST users.
\begin{figure}
\centering
\includegraphics{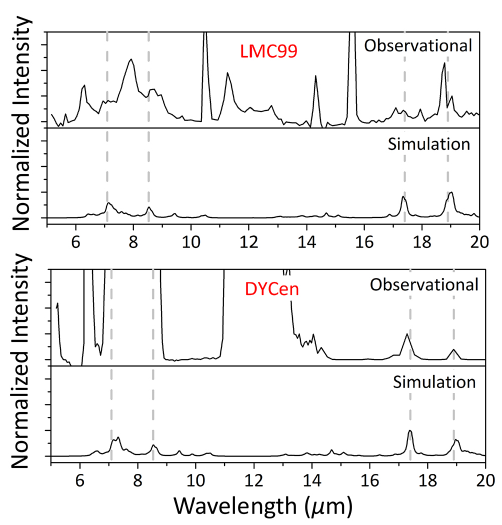}
\caption{Comparison of the continuum-subtracted {\it Spitzer Space Telescope} (top) and representative metallofullerenes simulated IR spectra (bottom) for two fullerene-rich objects showing very different C$_{60}$ 17.4$\mu$m/18.9$\mu$m band ratios and chemical abundances: the PN LMC 99 and the RCB star DY Cen. The guidelines show the features of pristine C$_{60}$ at 7.09, 8.53, 17.34 and 18.91 $\mu$m, according to our calculations. The representative metallofullerenes simulated spectra correspond to the simulations by \citet{barzaga2023}, which consider the metal abundances, ionization level, and endo/exo concentration \citep[see][for more details]{barzaga2023} and reproduce the observed 17.4$\mu$m/18.9$\mu$m band ratios. The intensity has been normalized according to the 17.4$\mu$m/18.9$\mu$m band ratio under the assumption that metallofullerenes are the only emitters causing the 17.4$\mu$m/18.9$\mu$m anomaly in the spectra; i.e., the more intense feature of the 17.4$\mu$m/18.9$\mu$m pair has been used to establish the reference in the normalization.
\label{fig:f7}}
\end{figure}

The individual metallofullerene spectra provide the several distinctive IR features of these complex metal-organic species. The metallofullerene mid-IR spectra (which, can be treated as unperturbed spectra) can be used to simulate different circumstellar and interstellar environments. The goal of the current data is to permit a precise modelling of the IR emission by metallofullerene species. For this, we provide also all the individual IR line positions accompanied by their corresponding intensities. Such data can be used to build the IR spectra representative of the metallofullerene species in several types of astronomical environments as in previous computational protocols \citep[see e.g.,][]{barzaga2023}. For example, depending on the thermal distribution of metallofullerenes in the circumstellar environments (Maxwell-Boltzmann distribution). Furthermore, the specific chemical abundances of the astrophysical environments, if known from observations or nucleosynthesis theoretical predictions, can be included in the simulation of the representative IR spectra of the metallofullerene species; e.g., as previously done by \citet{barzaga2023} and shown in Fig. 7 (see below), where the chemical abundances from nucleosynthesis models and observations where used, respectively, to simulate the metallofullerenes representative spectra in the circumstellar environments of PNe and RCB stars with fullerenes.

In \citet{barzaga2023} we considered the metal abundances, possible metallofullerenes formation reactions, ionization level, and endo/exo concentration to reproduce the large range of C$_{60}$ 17.4$\mu$m/18.9$\mu$m band ratios observed in very different fullerene-rich circumstellar environments like PNe and RCB stars. The simulated representative IR spectra of the metallofullerene species in two fullerene-rich objects (the PN LMC 99 and the RCB star DY Cen) are shown (for the first time) in Fig. 7 in comparison with their {\it Spitzer Space Telescope} spectra. Although the {\it Spitzer} IR spectra have a lower spectral resolution and sensitivity than the JWST ones, Fig. 7 shows that metallofullerenes have a significant contribution to the IR bands of neutral C$_{60}$ in circumstellar environments showing very different C$_{60}$ 17.4$\mu$m/18.9$\mu$m band ratios. The representative metallofullerenes
simulated IR spectra are consistent with the {\it Spitzer} observations, even assuming that metallofullerenes are the only emitters causing the 17.4$\mu$m/18.9$\mu$m anomaly in the spectra. From this exercise, we conclude that the metallofullerenes species expected to be present in space would depend on the specific physico-chemical properties of the circumstellar environment (i.e., metal abundances, formation reactions, ionization level, and endo/exo concentration) but the more prevalent metallofullerene species are generally those with the most abundant metals like Mg and Fe \citep[see also][]{barzaga2023}.

We remark that the individual spectra of all the metallofullerene species considered here display IR emission close to the four strongest IR features of neutral C$_{60}$ as well as a series of discrete features in the 6-9 $\mu$m spectral region, which can be of similar strength (exo species) or stronger (endo species) than the pristine C$_{60}$ features (see Figs.1$-$4). The rest of IR features, more specific to each metallofullerene species, are generally weaker. A detailed comparison of the 28 individual metallofullerene mid-IR simulated spectra with the available {\it Spitzer} spectral data of fullerene-rich sources is out of the scope this paper. For example, PNe with fullerenes display the general presence of a broad and complex (with multiple components/peaks and blends) IR feature at $\sim$6$-$9 $\mu$m \citep[e.g.,][]{Garcia2012,bernard-salas2012,sloan2014} that would be consistent (after intensity scaling) with most of the individual metallofullerene simulated spectra \citep[e.g.,][]{GaoLei23} but not enough to claim their detection/identification. In addition, such comparisons would not consider the metal abundances, possible metallofullerenes formation reactions, ionization level, and endo/exo concentration in the circumstellar environments around PNe (see above).

In short, the quantum chemistry-simulated metallofullerenes IR spectra presented here can be combined with astrophysical models, and their comparison with real JWST data could eventually lead to the detection of these complex metal-organic species in space.


\section{Acknowledgments}
We acknowledge support from the ACIISI, Gobierno de Canarias, and the European Regional Development Fund (ERDF) under a grant with reference PROID2020010051 as well as the State Research Agency (AEI) of the Spanish Ministry of Science and Innovation (MICINN) under grants PID2020-115758GB-I00 and PID2019-110091GB-I00 and the 'Mar\'ia de Maeztu' (CEX2018-000805-M) Program for Centers of Excellence in R\&D. This article is based upon work from COST Action NanoSpace, CA21126, supported by COST (European Cooperation in Science and Technology).We acknowledge the generous allocation of computer time at the Centro de Computaci\'on Cient\'{\i}fica at the Universidad Aut\'onoma de Madrid (CCC-UAM) and LaPalma Supercomputer at the Instituto de Astrof\'{\i}sica de Canarias.

\bibliography{Metallofullerenes_ApJS_dagh}{}
\bibliographystyle{aasjournal}



\end{document}